\documentclass[final,times,twocolumn,sort&compress]{elsarticle}
\usepackage{amsmath}
\usepackage{amssymb}
\usepackage{graphics}
\usepackage{epstopdf}

\journal{Journal of Magnetism and Magnetic Materials}

\begin{document}

\begin{frontmatter}

\title{Influence of a spatial anisotropy on presence of the intermediate one-half magnetization plateau of a spin-1/2 Ising-Heisenberg branched chain\tnoteref{grant}} 
\author[upjs]{Jozef Stre\v{c}ka\corref{coraut}}  
\cortext[coraut]{Corresponding author}
\ead{jozef.strecka@upjs.sk}
\author[upjs]{Katar\'ina Karl'ov\'a}  
\author[upjs]{Azadeh Ghannadan}  
\tnotetext[grant]{This work was supported under the grant Nos. VEGA~1/0531/19 and APVV-16-0186.}
\address[upjs]{Department of Theoretical Physics and Astrophysics, Faculty of Science, P. J. \v{S}af\'{a}rik University, Park Angelinum 9, 040 01 Ko\v{s}ice, Slovakia}

\begin{abstract}
A spin-1/2 Ising-Heisenberg branched chain constituted by regularly alternating Ising spins and Heisenberg dimers involving an additional side branching is exactly solved in a magnetic field by the transfer-matrix method. The spin-1/2 Ising-Heisenberg branched chain involves two different Ising and one Heisenberg coupling constants. The overall ground-state phase diagram is formed by three different ground states emergent depending on a mutual interplay between the magnetic field and three considered coupling constants: the modulated quantum antiferromagnetic phase, the quantum ferrimagnetic phase, and the classical ferromagnetic phase. It is shown that the interaction anisotropy connected to two different Ising coupling constants substantially influences a breakdown of the intermediate one-half magnetization plateau, which represents a macroscopic manifestation of the quantum ferrimagnetic phase. 
\end{abstract}

\begin{keyword}
Ising-Heisenberg branched chain, magnetization plateau, interaction anisotropy 
\end{keyword}
\end{frontmatter}

\section{Introduction}
Quantum Heisenberg spin chains provide an intriguing platform for an investigation of quantum magnetism using fully rigorous calculation methods being completely free of any uncontrolled approximation \cite{mat93}. A few exactly solved Ising-Heisenberg and Heisenberg branched spin chains have recently attracted considerable attention, because they may exhibit striking quantum critical points of Kosterlitz-Thouless and Gaussian type \cite{ver19,sou20,kar19}. Moreover, the Ising-Heisenberg and Heisenberg branched spin chains are not only purely theoretical models, but they are closely related to a few real-world experimental realizations from the family of polymeric coordination compounds \cite{wan10,kan10}. 
In the present work we will investigate a spin-1/2 Ising-Heisenberg branched chain whose magnetic structure is inspired by the heterobimetallic coordination polymer [(Tp)$_2$Fe$_2$(CN)$_6$(OCH$_3$)(bap)Cu$_2$(CH$_3$OH)$\cdot$2CH$_3$OH] (Tp=tris(pyrazolyl)hydroborate, bapH = 1,3-bis(amino)-2-propanol) \cite{kan10} to be further abbreviated as [Fe$_{2}$Cu$_{2}$]$_{\infty}$.

\section{Model and method}
\label{method}

\begin{figure}[t]
\centering
\includegraphics[scale=0.35]{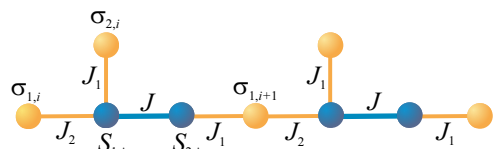}
\vspace{-0.4cm}
\caption{A schematic illustration of the spin-1/2 Ising-Heisenberg branched chain: orange (blue) circles denote lattice positions of the Ising (Heisenberg) spins.}
\label{fig1}
\end{figure}

The spin-1/2 Ising-Heisenberg branched chain, which is inspired by magnetic structure of the heterometallic coordination polymer [Fe$_{2}$Cu$_{2}$]$_{\infty}$ originally reported in Ref.~\cite{kan10}, can be defined through the Hamiltonian:
\begin{eqnarray} 
\hat{\cal{H}} \!\!\!&=&\!\!\! \displaystyle \sum_{i=1}^{N} \Bigl\{J \hat{\textbf{S}}_{1,i} \cdot \hat{\textbf{S}}_{2,i} +J_1\left(\hat{S}_{1,i}^z\hat{\sigma}_{2,i}^z+\hat{S}_{2,i}^z\hat{\sigma}_{1,i+1}^z\right) \nonumber \\ 
\!\!\!&+&\!\!\! J_2\hat{S}_{1,i}^z\hat{\sigma}_{1,i}^z- h\left(\hat{S}_{1,i}^z+\hat{S}_{2,i}^z + \hat{\sigma}_{1,i}^z + \hat{\sigma}_{2,i}^z \right) \Bigr\}.
\label{ham}
\end{eqnarray}
Here, $\hat{\sigma}_i^z$ and $\hat{S}_{i}^{\alpha}$ ($\alpha=x,y,z$) are the spin-1/2 operators ascribed to the Ising and Heisenberg spins, which are schematically shown in Fig. \ref{fig1} as orange and blue circles, respectively. This schematic illustration additionally involves also notation for three considered coupling constants: the coupling constant $J>0$ stands for the antiferromagnetic Heisenberg exchange interaction within the dimeric units from a backbone of the polymeric chain, while the coupling constants $J_1$ and $J_{2}$ correspond to two different Ising-type exchange interactions between the nearest-neighbor Ising and Heisenberg spins. Finally, the Zeeman's term $h$ accounts for a magnetostatic energy of the Ising and Heisenberg spins in a magnetic field, $N$ is the total number of unit cells and the periodic boundary condition $\sigma_{1,N+1}\equiv\sigma_{1,1}$ is imposed for simplicity. For further convenience, the total Hamiltonian \eqref{ham} can be rewritten as a sum of the cell Hamiltonians $\hat{\cal{H}}= \sum_{i=1}^N \hat{\cal{H}}_i$, where each cell Hamiltonian $\hat{\cal{H}}_i$ is defined by: 
\begin{eqnarray}  
\hat{\cal{H}}_{i} \!\!\!&=&\!\!\! J\hat{\textbf{S}}_{1,i} \cdot \hat{\textbf{S}}_{2,i}+ J_1\left(\hat{S}_{1,i}^z\hat{\sigma}_{2,i}^z+\hat{S}_{2,i}^z\hat{\sigma}_{1,i+1}^z\right) +J_{2} \hat{S}_{1,i}^z\hat{\sigma}_{1,i}^z \nonumber \\ 
\!\!\!&-&\!\!\! h\left(\hat{S}_{1,i}^z+\hat{S}_{2,i}^z + \hat{\sigma}_{2,i}^z\right) -\frac{h}{2}(\hat{\sigma}_{1,i}^z+\hat{\sigma_{1,i+1}}). 
\label{hamcell}
\end{eqnarray} 
The cell Hamiltonians $\hat{\cal{H}}_i$ obviously commute, i.e. $[\hat{\cal{H}}_i,\hat{\cal{H}}_j]=0$, which means that the partition function of the spin-1/2  Ising-Heisenberg branched chain can be partially factorized into the following product:
\begin{equation} 
\mathcal{Z} =\sum_{\{\sigma_{1,i}^z\}}\prod_{i=1}^N \sum_{\sigma_{2,i}^z}{\rm{Tr}}_{[S_{1,i}, S_{2,i}]}{\rm{e}}^{-\beta \hat{\mathcal{H}}_i}=\sum_{\{\sigma_{1,i}^z\}}\prod_{i=1}^N {\rm{T}}(\sigma_{1,i}^z,\sigma_{1,i+1}^z), \nonumber
\end{equation} 
where $\beta=1/(k_{\rm{B}}T)$, $k_{\rm{B}}$ is the Boltzmann's factor, $T$ is the absolute temperature, ${\rm{Tr}}_{[S_{1,i}, S_{2,i}]}$ denotes a trace over degrees of freedom of the Heisenberg dimer from the $i$-th unit cell and $\sum_{\{\sigma_{1,i}^z\}}$ denotes a summation over all possible configurations of the Ising spins from a backbone of the branched chain. The expression ${\rm{T}}(\sigma_{1,i}^z,\sigma_{1;i+1}^z) = {\rm{Tr}}_{[S_{1,i}, S_{2,i}]}{\rm{e}}^{-\beta \hat{\mathcal{H}}_i}$ is the usual transfer matrix obtained after tracing out spin degrees of freedom of two Heisenberg spins and the Ising spin $\sigma_{2,i}$ from the $i$-th unit cell. To proceed further with the calculations, we have to calculate eigenvalues of the cell Hamiltonian \eqref{hamcell} by performing a straightforward diagonalization in the local basis of the Heisenberg spins from the $i$-th unit cell:   
\begin{align}   
\begin{split} 
E_{i1,i2}&= \frac{J}{4} \pm \frac{J_1}{2}\left(\sigma_{1,i}^z+ \sigma_{1,i+1}^z\right)+\frac{J_2}{2}\sigma_{1,i}^z \mp h_1,\\
E_{i3,i4}&= -\frac{J}{4} \pm \frac{1}{2}\sqrt{\left[\frac{J_1}{2}(\sigma_{2,i}^z+\sigma_{1,i+1}^z)+\frac{J_2}{2}\sigma_{1,i}^z\right]^2 + J^2}, \nonumber
\end{split} 
\end{align}  
which should be shifted  by the field term $-\frac{h}{2}\left(\sigma_{1,i}^z+ \sigma_{1,i+1}^z\right)-h\sigma_{2,i}^z$ accounting for Zeeman's energy of the Ising spins.
The corresponding eigenvectors read
\begin{eqnarray} 
\label{vlastnevectorky}
|\varphi_{i1}\rangle \!\!\!&=&\!\!\! |\!\uparrow\rangle_{1,i}|\!\uparrow\rangle_{2,i}, \hspace{0.5cm} |\varphi_{i2}\rangle=|\!\downarrow\rangle_{1,i}|\!\downarrow\rangle_{2,i}, 
\nonumber \\
|\varphi_{i3}\rangle \!\!\!&=&\!\!\! c_+|\!\uparrow\rangle_{1,i}|\!\downarrow\rangle_{2,i} + c_-|\!\uparrow\rangle_{1,i}|\!\downarrow\rangle_{2,i}, \nonumber \\
|\varphi_{i4}\rangle \!\!\!&=&\!\!\! c_+|\!\uparrow\rangle_{1,i}|\!\downarrow\rangle_{2,i} - c_-|\!\uparrow\rangle_{1,i}|\!\downarrow\rangle_{2,i},
\end{eqnarray}
where 
\begin{eqnarray} 
\label{vlastnekoef}
c_{\pm}=\frac{1}{\sqrt{2}}\sqrt{1\pm\frac{J_1\left(\sigma_{1,i+1}^z-\sigma_{2,i}^z\right)+J_2\sigma_{1,i}^z}{\sqrt{\left[J_1(\sigma_{1,i+1}^z-\sigma_{2,i}^z)+J_2\sigma_{1,i}^z\right]^2+J^2}}}.
\end{eqnarray}
In this way one gets an explicit expression for the transfer matrix ${\rm{T}}(\sigma_{1,i}^z,\sigma_{1;i+1}^z)$:
\begin{align} 
\begin{split}
{\rm{T}}(\sigma_{1,i}^z&,\sigma_{1;i+1}^z)=\sum_{\sigma_{2,i}^z=\pm \frac{1}{2}}\sum_{j=1}^4{\rm{e}}^{-\beta E_{ji}}= 2{\rm{e}}^{\frac{\beta h}{2}\left(\sigma_{1,i}^z+\sigma_{1,i+1}^z\right)-\frac{\beta J}{4}} \\
&\times \left\{{\rm{e}}^\frac{\beta h}{2}\cosh\left[\frac{\beta}{2}\left(J_2\sigma_{1,i}^z+J_1\sigma_{1,i+1}^z +\frac{J_1}{2}+2h_1\right)\right] \right.\\ 
& + {\rm{e}}^{-\frac{\beta h}{2}}\cosh\left[\frac{\beta}{2}\left(J_2\sigma_{1,i}^z+J_1\sigma_{1,i+1}^z -\frac{J_1}{2}+2h_1\right)\right] \\
& +{\rm{e}}^{\frac{\beta J}{2}+\frac{\beta h}{2}}\cosh \left[\frac{\beta}{2}\sqrt{\left(J_2\sigma_{1,i}^z\!-\!J_1\sigma_{1,i+1}^z +\frac{J_1}{2}\right)^2 \!\!+ J^2}\right]  \\
& +\left.{\rm{e}}^{\frac{\beta J}{2}-\frac{\beta h}{2}}\cosh \left[\frac{\beta}{2}\sqrt{\left(J_2\sigma_{1,i}^z\!-\!J_1\sigma_{1,i+1}^z -\frac{J_1}{2}\right)^2 \!\!+ J^2}\right] \right\}. 
\label{Tranmat}
\end{split}
\end{align}
\begin{figure*}[t]
\centering
\hspace{-1cm}
\includegraphics[width=0.31\textwidth]{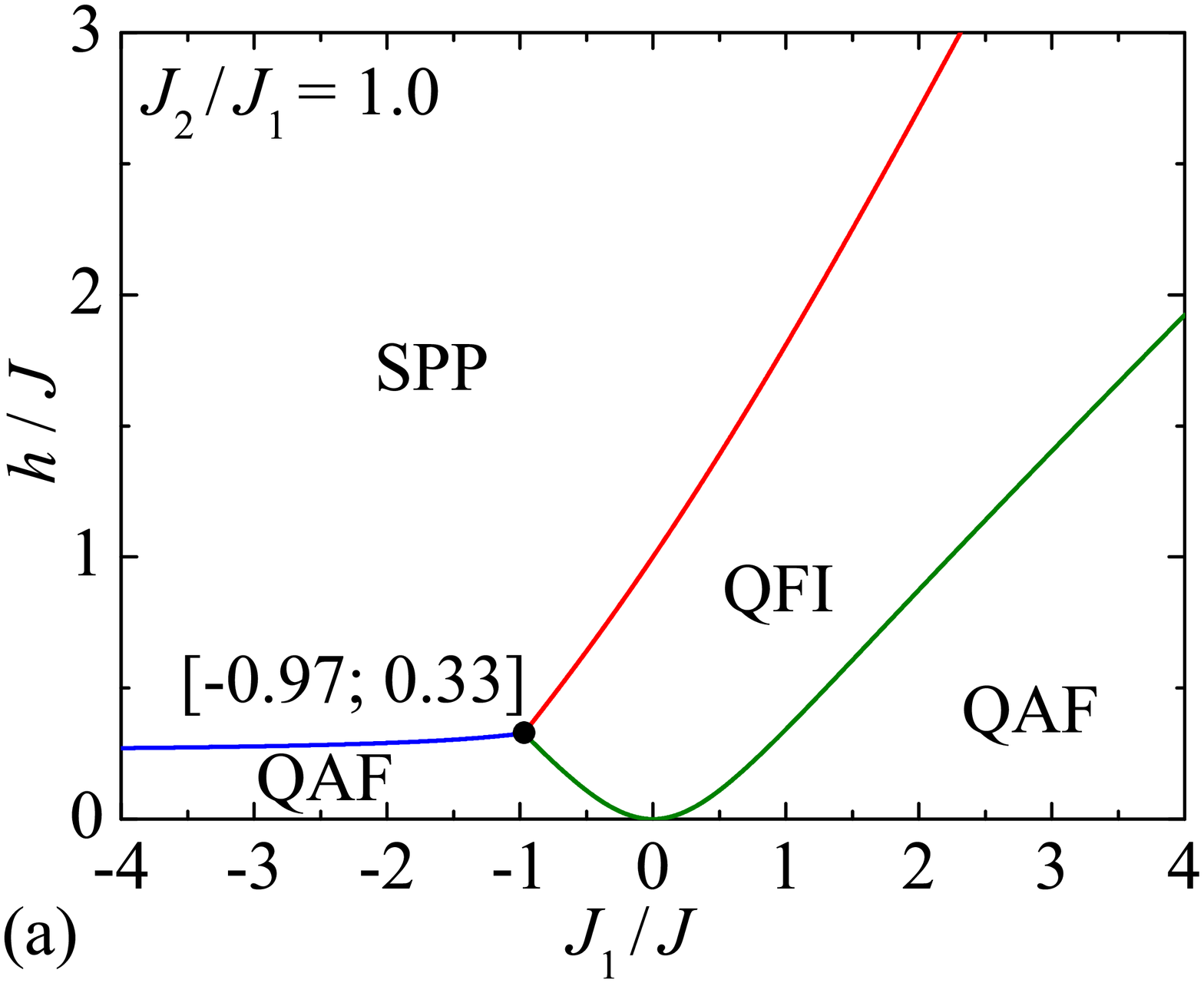}
\hspace{-1.2cm}
\includegraphics[width=0.31\textwidth]{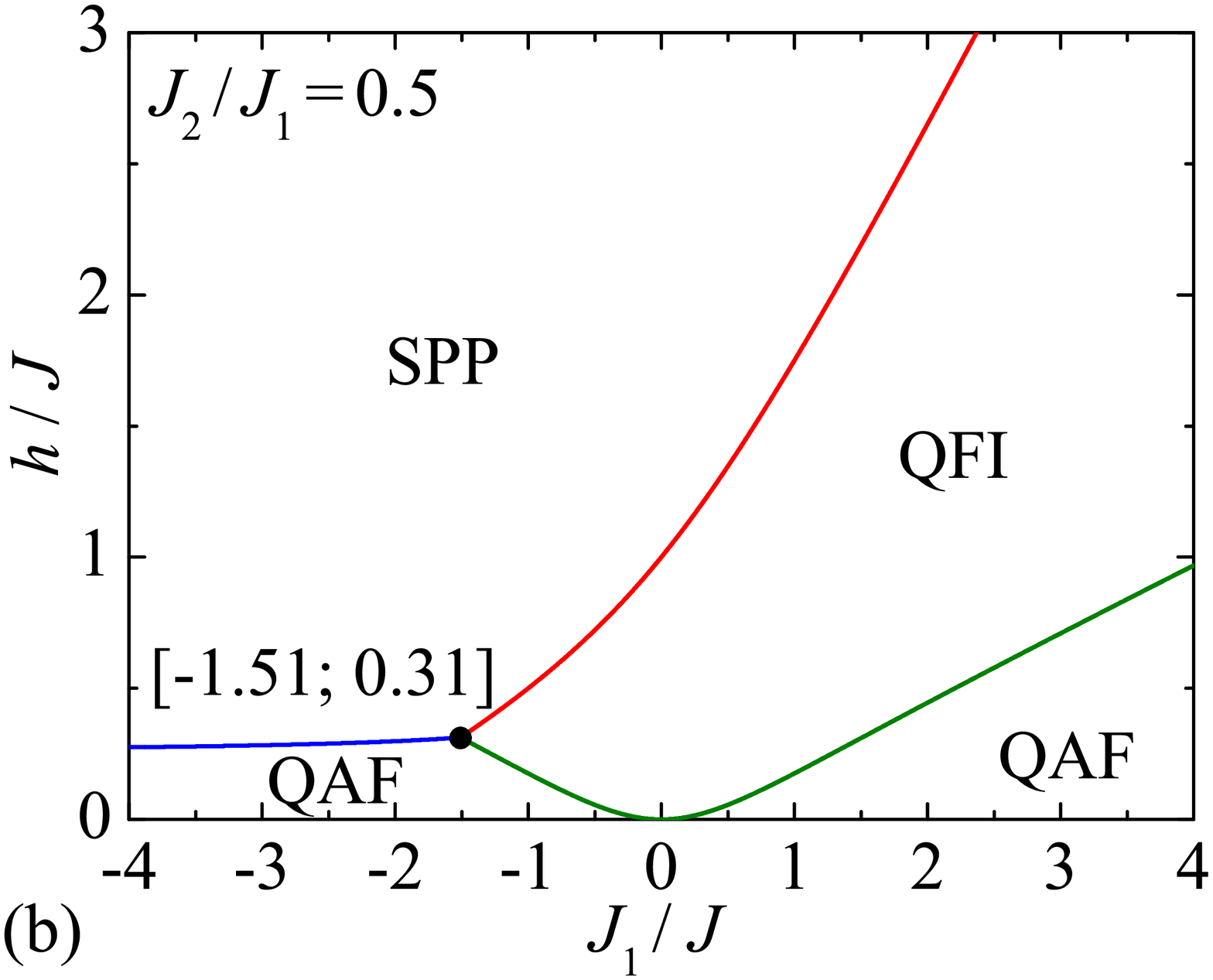}
\hspace{-1.2cm}
\includegraphics[width=0.31\textwidth]{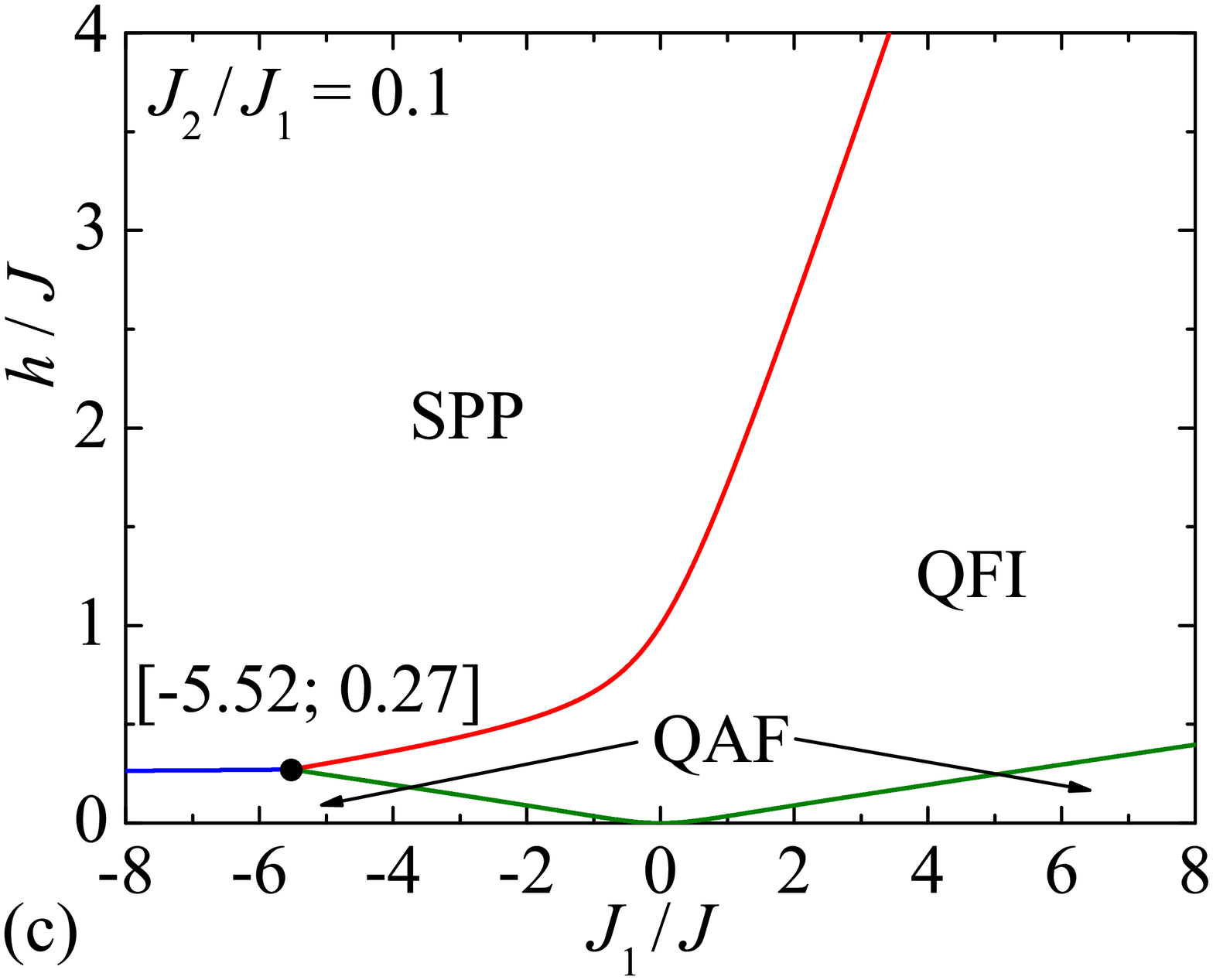}
\hspace{-1.2cm}
\includegraphics[width=0.31\textwidth]{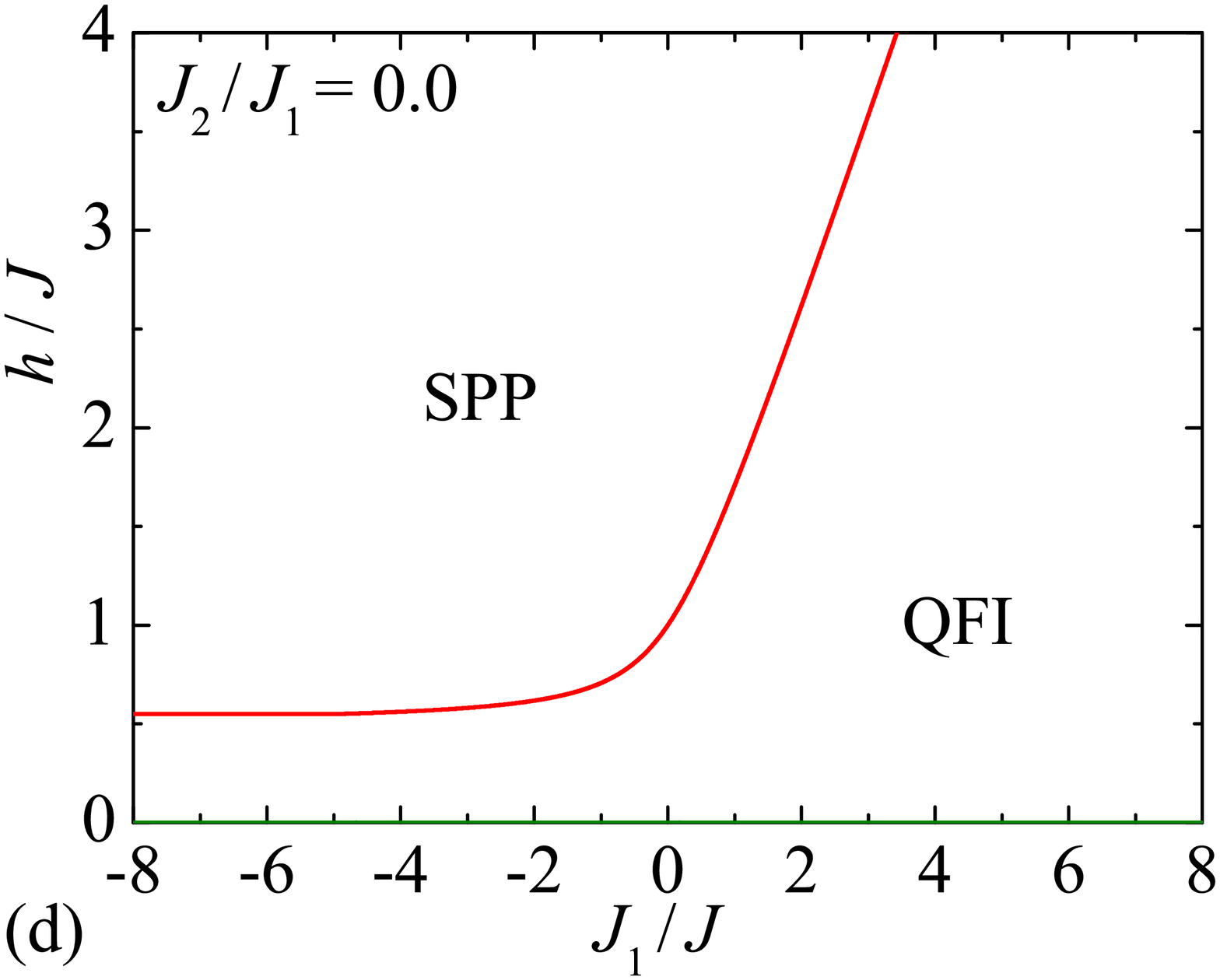}
\hspace{-1.2cm}
\vspace{-0.5cm}
\caption{Ground-state phase diagrams of the spin-1/2 Ising-Heisenberg branched chain in the $J_1/J-h/J$ plane for four different values of the interaction ratio: 
(a) $J_2/J_1 = 1.0$; (b) $J_2/J_1 = 0.5$; (c) $J_2/J_1 = 0.1$; (d) $J_2/J_1 = 0.0$. The triple-point coordinates are given in square brackets.}
\label{fig2}
\end{figure*}
The partition function of the spin-1/2 Ising-Heisenberg branched chain can be expressed in terms of two eigenvalues $\lambda_{+}$ and $\lambda_{-}$ of the transfer matrix (\ref{Tranmat}):
\begin{eqnarray} 
\label{Zfinal}
{\cal{Z}}=\sum_{\{\sigma_{1,i}\}}\prod_{i=1}^N {\rm{T}}\,(\sigma_{1,i}^z,\sigma_{1;i+1}^z) =  {\rm{Tr}}~{\rm{T}}^N=\lambda_+^N+\lambda_-^N,
\end{eqnarray}
which can be written in this compact form:
\begin{eqnarray} 
\label{lambdy}
\lambda_{\pm}=\frac{1}{2} \left[ T_1+T_2 \pm \sqrt{\left(T_1-T_2\right)^2+4T_3T_4} \right].
\end{eqnarray}
The parameters $T_i$ ($i=1-4$) mark four elements of two-by-two transfer matrix (\ref{Tranmat}):
\begin{eqnarray} 
T_1={\rm{T}}(+,+),\,\, T_2={\rm{T}}(-,-),\,\, T_3={\rm{T}}(+,-),\,\, T_4={\rm{T}}(-,+), \nonumber
\end{eqnarray}
which correspond to four possible states of the Ising spins $\sigma_{1,i}$ and $\sigma_{1,i+1}$ ($\pm$ applies for $\sigma = \pm 1/2$). In thermodynamic limit $N\rightarrow \infty$ the Gibbs free energy can be expressed through larger eigenvalue of the transfer matrix: 
\begin{eqnarray} 
\label{G}
G=-k_{\rm{B}}T\lim_{N\rightarrow\infty}\frac{1}{N}\ln{\cal{Z}}=-k_{\rm{B}}T\ln\lambda_+.
\end{eqnarray}
Other quantities can be subsequently derived from the Gibbs free energy \eqref{G} using standard relations. 

\section{Results and discussion}

\begin{figure*}[t]
\centering
\hspace{-1cm}
\includegraphics[width=0.33\textwidth]{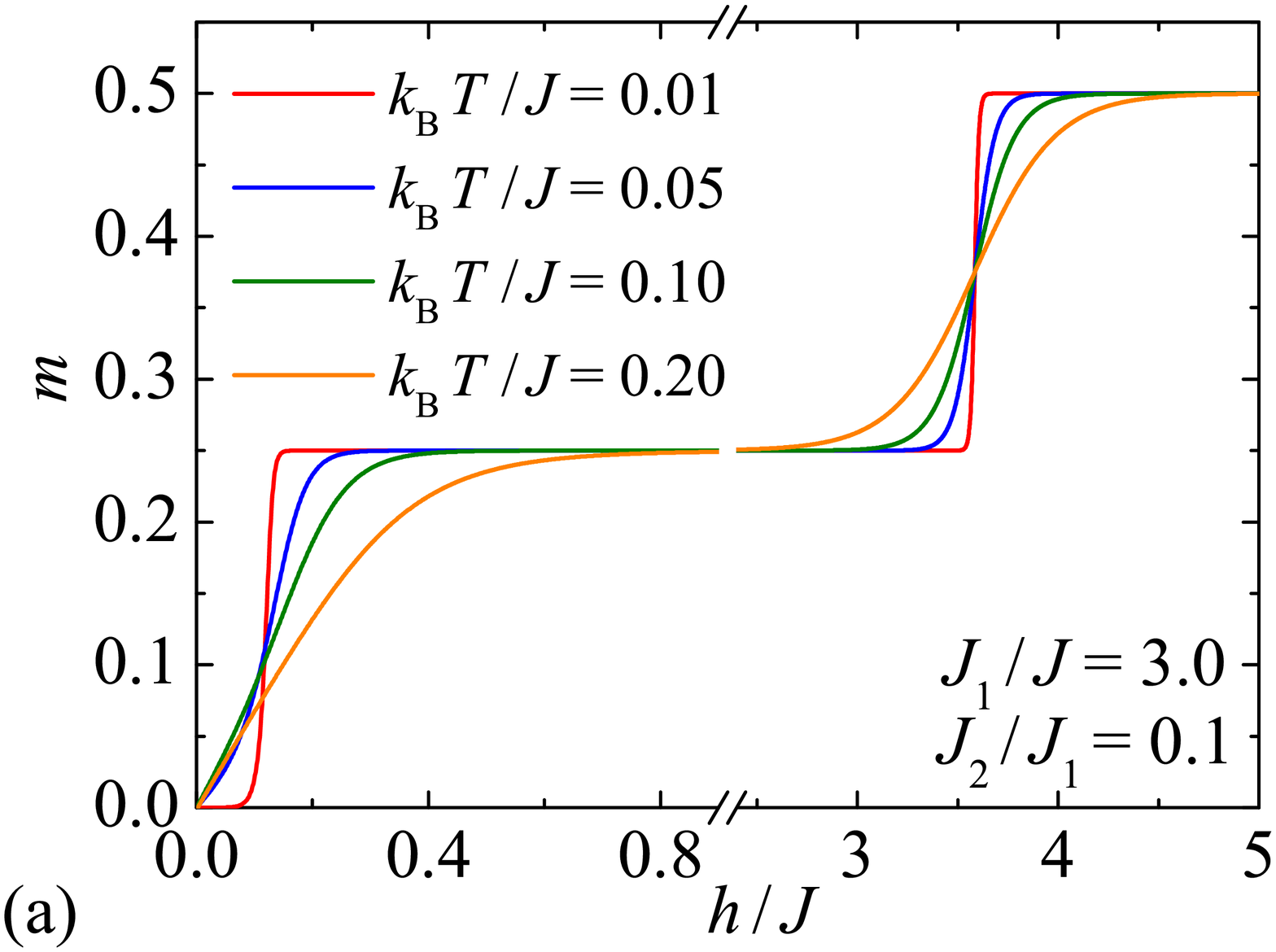}
\hspace{-0.6cm}
\includegraphics[width=0.33\textwidth]{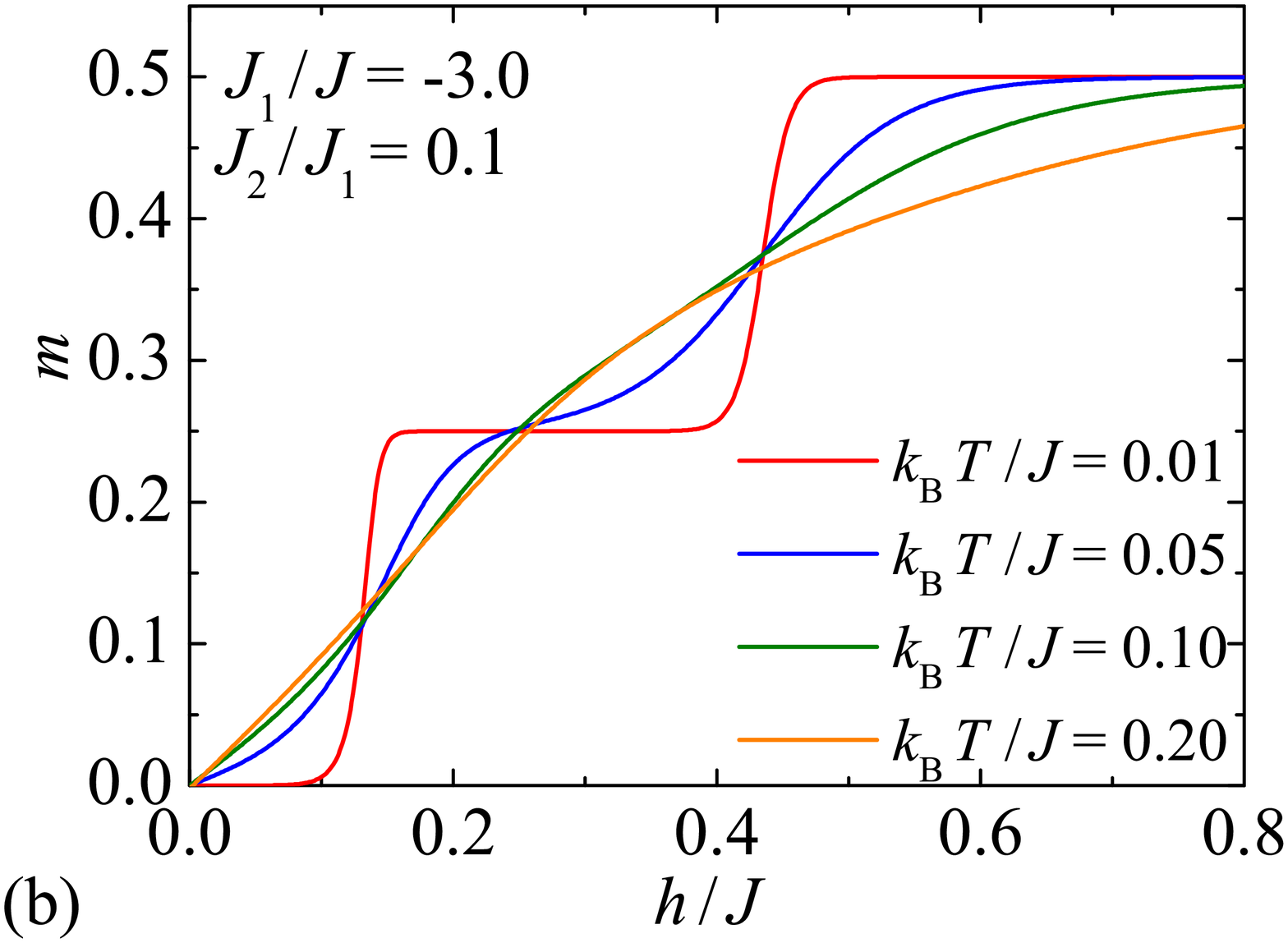}
\hspace{-0.6cm}
\includegraphics[width=0.33\textwidth]{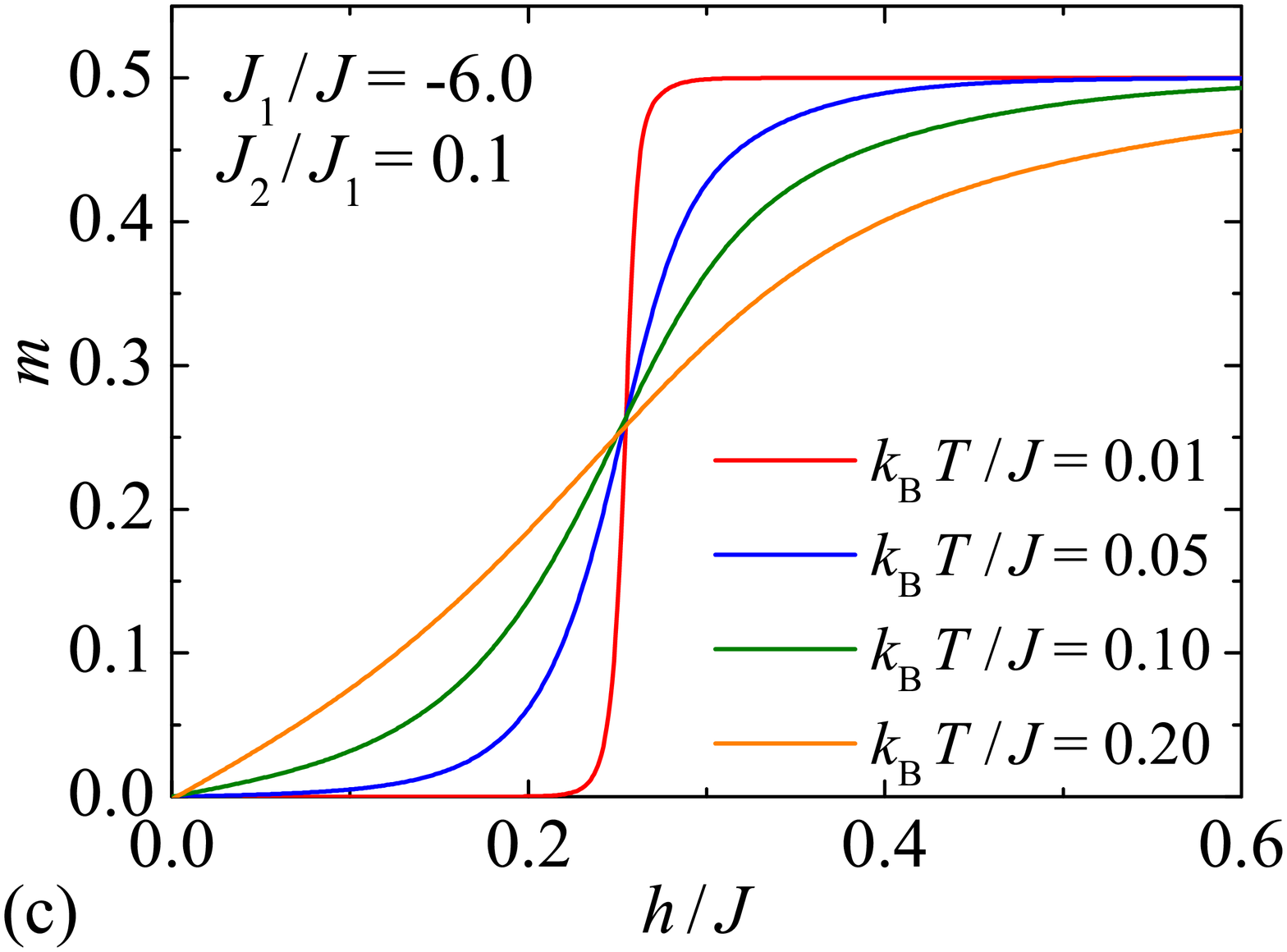}
\hspace{-1cm}
\vspace{-0.5cm}
\caption{A few typical isothermal magnetization curves of the spin-1/2 Ising-Heisenberg branched chain for the fixed value of the interaction anisotropy $J_2/J_1 = 0.1$ and three different values of the interaction ratio: (a) $J_1/J = 3.0$; (b) $J_1/J = -3.0$; (c) $J_1/J = -6.0$.}
\label{fig3}
\end{figure*}

Let us begin discussion of the most interesting results by a comprehensive analysis of the ground state. It turns out that the spin-1/2 Ising-Heisenberg branched chain may exhibit  just three different ground states referred to as the quantum antiferromagnetic phase (QAF):
\begin{eqnarray}
|{\rm{QAF}}\rangle \!\!\!\!&=&\!\!\!\! \prod_{i=1}^{N/2}\!\left\{|\!\uparrow\rangle_{\sigma_{1,2i-1}}|\!\uparrow\rangle_{\sigma_{2,2i-1}} \!\left(a_+|\!\downarrow,\!\uparrow\rangle
\!-\!a_-|\!\uparrow,\!\downarrow\rangle\right)_{S_{1,2i-1}, S_{2,2i-1}} \right.\nonumber \\
\!\!\!\!&\otimes&\!\!\!\! \left. \!|\!\downarrow\rangle_{\sigma_{1,2i}}|\!\downarrow\rangle_{\sigma_{2,2i}} \left(a_-|\!\downarrow,\!\uparrow\rangle\!-\!a_+|\!\uparrow,\!\downarrow\rangle\right)_{S_{1,2i},S_{2,2i}}\!\right\}\!, \nonumber 
\label{qaf}
\end{eqnarray}
the quantum ferrimagnetic phase (QFI):
\begin{eqnarray}
{|\rm{QFI}}\rangle = \prod_{i=1}^{N} |\!\uparrow\rangle_{\sigma_{1,i}}|\!\uparrow\rangle_{\sigma_{2,i}} \left(b_+|\!\downarrow,\!\uparrow\rangle\!-\!b_-|\!\uparrow,\!\downarrow\rangle\right)_{S_{1,i}, S_{2,i}},\nonumber
\label{qfi}
\end{eqnarray}
and the saturated paramagnetic phase (SPP): 
\begin{eqnarray}
|{\rm{SPP}}\rangle = \prod_{i=1}^{N}|\!\uparrow\rangle_{\sigma_{1,i}} |\!\uparrow\rangle_{\sigma_{2,i}}|\!\uparrow,\!\uparrow\rangle_{S_{1,i},S_{2,i}}. \nonumber
\label{spp}
\end{eqnarray}
For a shorthand notation the QAF and QFI ground states are defined through the probability amplitudes:
\begin{eqnarray}
a_{\pm} = \frac{1}{\sqrt{2}}\sqrt{1\pm\frac{J_1 + \frac{J_2}{2}}{\sqrt{\left(J_1 + \frac{J_2}{2}\right)^2 + J^2}}} 
\end{eqnarray}
and 
\begin{eqnarray}
b_{\pm} = \frac{1}{\sqrt{2}}\sqrt{1\pm\frac{\frac{J_2}{2}}{\sqrt{\left(\frac{J_2}{2}\right)^2 + J^2}}}. 
\end{eqnarray}
It is worth mentioning that the QAF ground state with translationally broken symmetry is consistent with existence of zero magnetization plateau in a zero-temperature magnetization curve due to a null total magnetization, while the QFI ground state is responsible for presence of the intermediate one-half plateau if the total magnetization is scaled with respect to its saturation value. 

The ground-state phase diagrams of the spin-1/2 Ising-Heisenberg branched chain are plotted in Fig. \ref{fig2} in the $J_1/J-h/J$ plane for four selected  values of the interaction anisotropy $J_2/J_1$. It can be concluded that the ground-state phase diagrams are formed regardless of the interaction anisotropy $J_2/J_1$ by the same three ground states QAF, QFI and SPP as previously reported in Ref. \cite{kar19} for the isotropic case with $J_2/J_1 = 1$, see Fig. \ref{fig2}(a). The interaction anisotropy, i.e. the decline of the interaction ratio from the value $J_2/J_1 =1$, merely causes an extension of the QFI ground state down to lower values of the interaction ratio $J_1/J$. On the other hand, the QAF ground state is gradually suppressed by the interaction anisotropy (i.e. when the interaction ratio $J_2/J_1$ decreases) until the QAF ground state completely disappears from the phase diagram in the limit $J_2/J_1 \to 0$.   

To verify the aforedescribed behavior, a few typical isothermal magnetization curves of the spin-1/2 Ising-Heisenberg branched chain are displayed in Fig. \ref{fig3} for the fixed value of the interaction anisotropy $J_2/J_1 = 0.1$ and three selected values of the interaction ratio $J_1/J = 3.0$, -3.0 and -6.0, respectively. It can be seen that a relatively wide one-half plateau and narrow zero plateau can be observed by considering the antiferromagnetic Ising coupling $J_1/J = 3.0$ [see Fig. \ref{fig3}(a)], while the width of zero plateau extends and of one-half plateau shrinks by considering the ferromagnetic Ising coupling $J_1/J = -3.0$ [see Fig. \ref{fig3}(b)]. If the ferromagnetic Ising interaction is sufficiently strong one detects a mere existence of zero plateau and a full breakdown of the one-half plateau (see Fig. \ref{fig3}(c) for $J_1/J = -6.0$). It is noteworthy that the depicted magnetization curves are in a perfect accordance with the established ground-state phase diagrams (c.f. Figs. \ref{fig2} and \ref{fig3}), whereby the intermediate one-half plateau is absent if a relative strength of the ferromagnetic Ising coupling constant exceeds the particular value ascribed to a triple coexistence point of the QAF, QFI and SPP ground states.  

\section{Conclusion}
\label{conclusion}

In the present paper we have exactly solved using the transfer-matrix method the spin-1/2 Ising-Heisenberg branched chain with two different Ising and one Heisenberg coupling constants in a magnetic field. It has been verified that the investigated quantum spin chain may exhibit just three different ground states QAF, QFI and SPP depending on a mutual interplay between the magnetic field and three considered coupling constants. The QAF and QFI ground states with a quantum entanglement between the Heisenberg dimers are responsible for presence of intermediate zero and one-half plateaus in zero- and low-temperature magnetization curves, whereby a relative size of the intermediate magnetization plateaus depends basically on the interaction anisotropy. A full breakdown of the intermediate one-half magnetization plateau has been additionally detected for the particular case with sufficiently strong ferromagnetic Ising coupling constants.

\end{document}